# A Scientific Trigger Unit for Space-Based Real-Time Gamma Ray Burst Detection

# I – Scientific Software Model and Simulations

Stéphane Schanne[1], Hervé Le Provost[2], Pierre Kestener[3], Aleksandra Gros[1], Marin Cortial[1], Diego Götz[1], Patrick Sizun[2], Frédéric Château[2], Bertrand Cordier[1]

*Abstract*–The on-board Scientific Trigger Unit (UTS) is designed to detect Gamma Ray Bursts (GRBs) in real-time, using the data produced by the ECLAIRs camera, foreseen to equip the future French-Chinese satellite mission SVOM (Space-based Variable Objects Monitor). The UTS produces GRB alerts, sent to the ground for GRB follow-up observations, and requests the spacecraft slew to repoint its narrow field instruments onto the GRB afterglow. Because of the diversity of GRBs in duration and variability, two simultaneously running GRB trigger algorithms are implemented in the UTS, the so called Image Trigger performing systematic sky image reconstruction on time scales above 20 s, and the Count-Rate Trigger, selecting a time scale from 10 ms to 20 s showing an excess in count-rate over background estimate, prior to imaging the excess for localization on the sky. This paper describes both trigger algorithms and their implementation in a library, compiled for the Scientific Software Model (SSM) running on standard Linux machines, and which can also be cross-compiled for the Data Processing Model (DPM), in order to have the same algorithms running on both platforms. While the DPM permits to validate the hardware concept and benchmark the algorithms (see paper II), the SSM allows to optimize the algorithms and estimate the GRB trigger-rate of ECLAIRs/UTS. The result of running on the SSM a dynamic photon by photon simulation based on the BATSE GRB catalog is presented.

## I. INTRODUCTION

### A. Gamma-Ray Bursts

Gamma-Ray Bursts (GRBs) are transient sources appearing randomly on the sky as short flashes in gamma-rays, lasting mostly between about 100 ms and a few minutes. They are believed to be produced by internal shocks in relativistic jets emitted just after the formation of stellar-mass black holes. Rapidly decaying afterglows in X-rays, the visible band, infra-red and radio follow the gamma-ray flash; they are believed to be produced by shocks when the jet encounters the circum-stellar medium. The afterglow observation is crucial to further characterize the GRB event, in particular to determine its distance via the measurement of its redshift in the visible and infra-red. Most GRBs appear to be events at cosmological distances, and as such are interesting for a large part of the astrophysical community. A variety of scientific studies are accessible through GRBs observations, in particular :

(i) physics of jet acceleration and relativistic shocks,

(ii) GRB phenomenology, long duration GRBs being probably linked to supernova events and short GRBs being believed to be related to compact object mergers,

(iii) astrophysics of GRB host galaxies and the study of the foreground universe illuminated by those cosmic light houses,

(iv) cosmology with the study of the star formation history of the Universe, the reionization phase of the early Universe, and measurements of the cosmological parameters,

(v) fundamental physics, such as Lorentz invariance tests, GRBs as prominent sources of cosmic rays, neutrinos and gravitational waves.

To be useful to the community, a GRB detection must be provided quickly (< minutes), and with accurate localization (< arcminutes).

### B. The SVOM mission

The SVOM mission (Space-Based Variable Objects Monitor) [1] dedicated to GRB studies, expected to operate by the end of the decade, is currently developed in French-Chinese cooperation. Its main scientific requirements are:

(i) detection of all know types of GRBs, especially high-z GRBs thanks to low energy threshold of its trigger (4 keV),

(ii) fast alerts sent to ground (within ~30 s) with good localization (~10 arcmin in γ-rays, sub-arcmin after ~5 min).

(iii) prompt-emission spectral and temporal monitoring in X/γ-rays (4 keV to a few MeV) and the visible band

(iv) afterglow identification and localization down to arcsec level, with redshift indicators,

(v) optimized ground follow-up by pointing the satellite mainly towards the night sky, allowing for a good fraction of redshift measurements.

The SVOM space segment consists of a 3-axis stabilized low Earth-orbit (~600 km) satellite with rapid repointing capability. The instrumental concept of SVOM is presented in Fig. 1. GRBs are first observed with the onboard X/γ-ray telescope ECLAIRs. The associated onboard processing electronics UTS (Unit for Triggering and Scientific processing), continuously processes the ECLAIRs data stream to detect and localize GRBs in near real-time. Once detected the UTS sends a GRB alert sequence to the ground via a dedicated VHF network and initiates a spacecraft reorientation maneuver. This places the GRB afterglow within minutes into to fields-of-view (adapted to the ECLAIRs localization error-box) of an X-ray telescope and a visible telescope. The later refine to sub-arcmin accuracy the GRB localization and send them also via VHF to the ground. A high fraction of the

SVOM GRBs will have redshifts measured by ground follow-up telescopes, since the pointing strategy ensures that GRBs are mostly observed towards the night side on Earth; it also avoids the galactic plane and bright sources like Sco X-1 to be present in the ECLAIRs FOV.

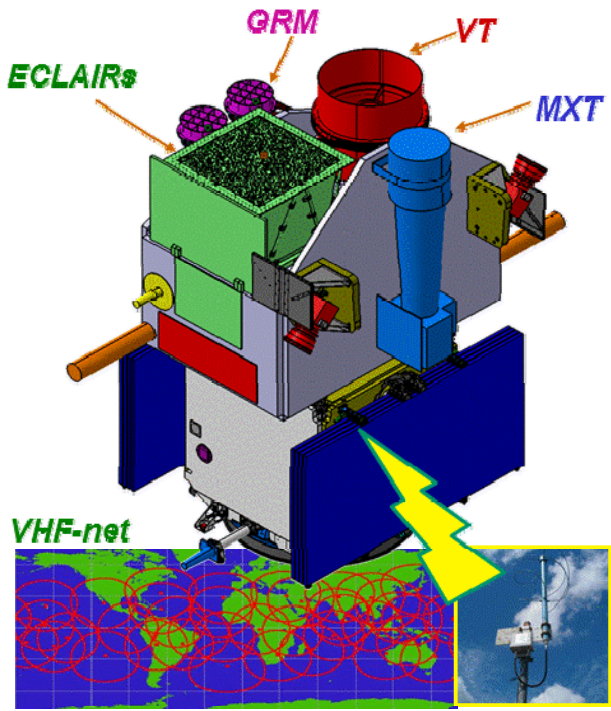

Fig. 1 – SVOM mission concept: a large portion of the night sky is monitored by two large field-of-view instruments ECLAIRs ("flash of lightening" in French) and GRM (Gamma-Ray Monitor). The onboard electronics UTS of ECLAIRs localizes GRBs and initiates spacecraft reorientation maneuvers to permit follow-up observations by the narrow field-of-view instruments VT (Visible Telescope) and MXT (Micro-channel X-ray Telescope). The UTS also transmits GRB alerts to ground via the VHF network to the dedicated GFTs (Ground Follow-up Telescopes) and GWAC (Ground Wide-Angle Camera) as well as the whole GRB observer community.

### C. The ECLAIRs telescope

The ECLAIRs γ-ray camera [2][3] is sensitive from 4 keV to more than 150 keV. It uses a square detector matrix with a 1024 cm$^2$ efficient area, made of 80×80 CdTe pixels of 4×4×1 mm$^3$ each, passively cooled to -20°C, arranged on a grid with 4.5 mm spacing. The detector is placed 46 cm under a 0.6 mm-thick Tantalum coded mask of 54×54 cm$^2$, arranged in a self-supporting random pattern of 100×100 pixels with a 40% opening fraction, which permits source localization on the sky with accuracy better than 10 arcmin. The 2 sr field-of-view is defined by a side-shield assembly, which also reduces the background on the detector. The residual background is mainly due to cosmic X-ray background photons (CXB) modulated by the passage of the Earth across the field-of-view each orbit, which results from the pointing strategy towards the night sky.

With its good sensitivity on a large portion of the sky, ECLAIRs is designed to efficiently observe GRBs. Thanks to its low energy threshold, ECLAIRs is particularly well suited to detect high redshift GRBs.

### D. The UTS onboard processing system

The UTS acquires in real-time all ECLAIRs camera hits (detector pixel, energy, time), runs the GRB trigger to detect and localize GRBs in near real-time, and generates GRB alerts over the VHF, and issues satellite repointing requests. With no permanent high bandwidth data-link to ground on a low-Earth-orbit satellite, these functions can only be performed by an onboard system.

The UTS is implemented as a digital processing system, based on FPGAs and CPUs from Atmel, which are radiation tolerant and not subject to export restrictions to China. The UTS hardware and the benchmarking tests of the software on the target CPU system are presented in a subsequent paper [4]. The foreseen algorithms were presented in [5]. The current paper is focused on the definition of the actually implemented algorithms and their scientific performances obtained on a specific GRB database.

## II. GRB DIVERSITY AND IMPACT ON THE TRIGGER

The diversity of the GRB prompt emission impacts on the scientific parameters of the UTS trigger algorithms.

GRB energy spectra can be hard or soft, X-ray rich, or with low Epeak, especially for high-redshift GRBs. Therefore to detect GRBs of all spectral types the UTS uses 4 overlapping energy strips, typically 4-50, 4-80, 4-120 keV and a more standard 15-50 keV band (Fig. 2). This choice maximizes the number of detected GRBs after imaging on a synthetic GRB database.

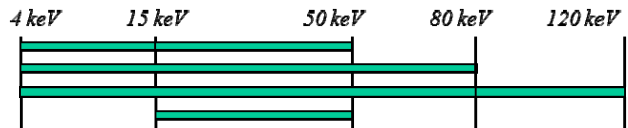

Fig. 2 – The 4 energy strips used by the UTS trigger algorithms.

GRB durations span from very short (10 ms) to very long (up to 20 min) and their light-curves can be smooth or spiky. Since the sky image reconstruction to search for a new source takes about 2 s on the target UTS hardware, systematic sky imaging on the smallest GRB time-scales is impossible.

It is systematically performed on long time scales only (>20 s) by the so-called "Image Trigger" algorithm. This algorithm aims at finding low-significance long-duration GRBs.

On shorter time-scales, the so-called "Count-Rate Trigger" algorithm first selects a time-window on which a count-rate excess is detected prior to imaging. For a quick response this algorithm also aims at triggering on first spikes in GRB light curves, instead of the whole GRB duration.

Interesting GRBs may actually be faint in γ-rays (especially the distant ones). Thus in the signal-to-noise ratio images (SNR = counts/sqrt(variance) for each sky pixel) where new sources are searched, in order to keep an acceptable false trigger rate, the threshold is chosen as low as possible (6.5σ, a value close to the theoretical limit and resulting from simulations with Earth presence in the field-of-veiw).

Since GRBs appear randomly with an isotropic distribution on the sky, most of them are detected about 30° off axis. To integrate less background, the Count-Rate Trigger analyzes thus 9 detector zones: additionally to the full detector, the 4 detector halves (upper, lower, left and right) and the 4 quadrants are considered (Fig. 3).

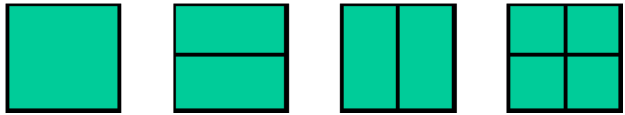

Fig. 3 – The 9 zones of the detection plane used by the UTS Count-Rate trigger algorithm.

## III. TRIGGER ALGORITHMS

### A. The Image Trigger

The Image **Trigger algorithm** is implemented as a cyclic process, running on the 4 energy strips every 20.48 s.

At each of these periods, first the shadowgram (detector pixel array of accumulated counts) is built. The background, mainly due to CXB photons with an isotropic distribution in the field-of-view has a convex shape and is ~60% of the time partially obscured by the Earth; thus it is spatially non-flat in the shadowgram. If not corrected on long time scales, this induces artifacts in the deconvolved sky images, which would require increasing the detection threshold with consequent loss of faint sources. To overcome this, the background shape is first modeled by fitting a 2D $2^{nd}$ degree polynomial function to the shadowgram. The sky reconstruction is performed on the background corrected count and variance shadowgrams, using a FFT-based balanced mask pattern deconvolution [6], which produces sky images in counts, variance and SNR (Fig. 4).

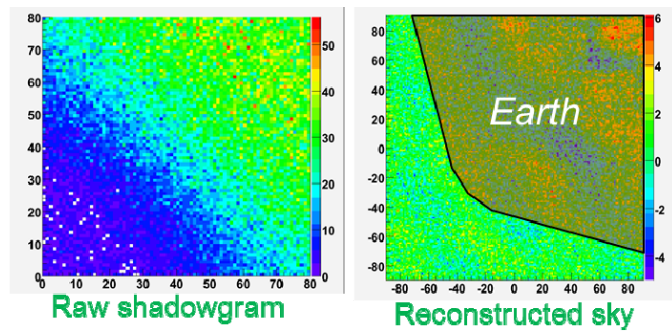

Fig. 4 – (left) Count shadowgram (80×80 detector pixels) showing the spatially non-flat background induced by CXG photons masked by the Earth present in the field-of-view. (right) Reconstructed sky image in SNR (number of sigmas, 200×200 sky pixels) after background correction. The zone of the sky masked by the Earth is shaded in gray, outside this zone no sky pixel exceeds the detection threshold.

Those sky images are put into a sky image history (Fig. 5), which constructs subsequent sky images on 7 time-scales ranging from 20.48 s to 1310.72 s (about 20 min), by summation of the count and variance sky images obtained at previous time-scale. The summed SNR image is then calculated from these images. For each time-scale (except the shortest time-scale of 20.48 s), a set of two sky images, half overlapping in time is constructed. The deconvolution being linear, this sky summation, although expensive in terms of memory need, is preferred to deconvolutions at all time-scales, which would be more demanding in terms of CPU power.

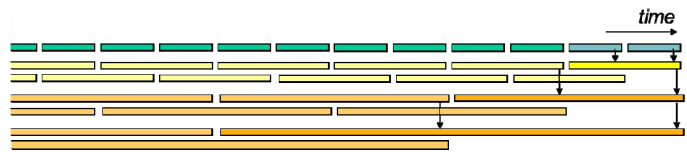

Fig. 5 - Principle of sky summation used to produce the sky history of the UTS Image Trigger (4 out of 7 time-scales shown here).

In parallel to the sky image history, a sky exposure history is built on the same time-scales, taking into account the Earth position vector, sampled each 20.48 s period.

### B. New source localization

All the sky images built by the Image Trigger algorithm at a given 20.48 s period, are subsequently processed by the **New-Source Localization** algorithm.

This algorithm searches for peaks in the SNR image exceeding threshold. Sky portions with too low exposure (obtained from the sky exposure history) are ignored. Peaks corresponding to positions referenced in the known source catalog are also ignored (this catalog is recomputed in local coordinates after each slew). The detection of a new source with SNR above threshold is a GRB trigger. The corresponding part of the count image is fit with the instruments point-spread function, a 2D Gaussian with the width fixed according to the mask pixel $m$ over detector pixel $d$ ratio $m/d$. The centroid of the Gaussian defines the source position. With an accuracy better than 10 arcmin (at 90% confidence level), it is used for the alert message on the VHF and the spacecraft reorientation request.

### C. The Count-Rate Trigger

The **Count-Rate Trigger** algorithm is implemented as a cyclic process running every 2.56 s in two subsequent steps.

In the first step, count-rate excesses are searched for the following parameter combinations: 4 energy strips, 9 detector zones and 12+11 time-scales ranging from 10 ms to 20.48 s (Fig. 6). For each combination, the total number of counts N is compared to the corresponding background-count estimate B, obtained from fitting a trend on previous counts N on long-time scales (which were not previously in excess). Count-excesses in SNR defined as (N-B)/sqrt(B) above a given threshold (chosen close to noise, to get an fair number of excesses) are stored in an excess buffer.

In the second step, the excess buffer is analyzed, to determine the best excess not yet processed and not too old. The shadowgram is built with photons from the excess time-scale and energy strip, extracted from a separate photon buffer. The counts and variance sky images are constructed and a **New-Source Localization** (similar to the one used for the Image Trigger) is launched.

Cyclic execution of both steps (Fig. 7) can produce subsequent VHF messages with refined source positions as the GRB evolves, until time-out or until the source is detected with confidence high enough to slew the spacecraft, at which point the trigger process is stopped, and reinitialized once the spacecraft attitude is again stabilized.

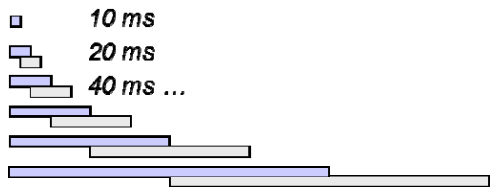

Fig. 6 – Time-scales considered in the Count-Rate trigger. From the shortest time-scale of 10 ms, 11 longer duration time-scales (5 shown here) are built by doubling the time interval at each step, and 11 more by considering overlapping time intervals. The counts on all those time-scales are computed by difference of the two appropriate values, read in a large circular memory buffer, fed with the integrated count values each 10 ms time-scale.

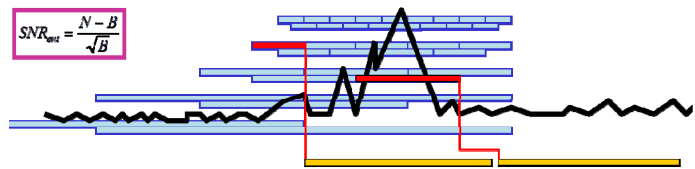

Fig. 7 – Illustration of the UTS Count-Rate Trigger. Multiple time scales are evaluated for count-rate increases. The best excess found at a cycle is imaged. At a later cycle, the new best excess not yet processed is imaged. VHF alert messages can be sent after each cycle.

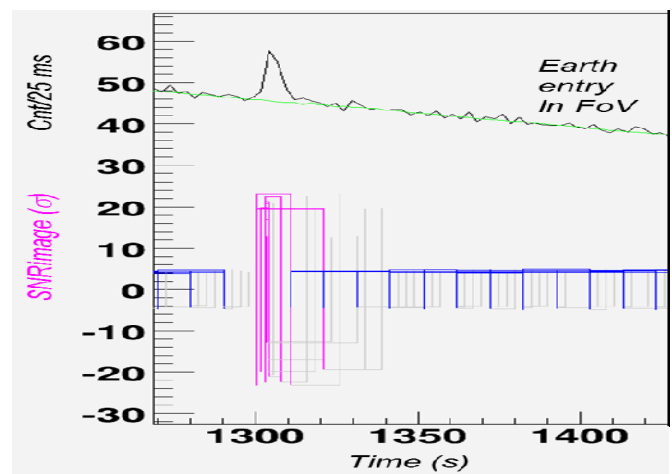

Fig. 8 – Example of a simulated GRB detection by the UTS Count-Rate Trigger. In black: count rate evolution with time; the Earth entry in the field-of-view results in a decrease of the count-rates. In green: fitted background estimate. In pink (blue): SNR significance level in the image (number of sigmas) of the best excesses exceeding (not exceeding) the threshold at a given 2.56 s period of the algorithm; the moment the image is analyzed is indicated by the gray vertical bar with the same significance level. In this example the best detection reaches 24 σ in an image covering a 10 s time-slice, obtained 7.5 s after the end of the time-slice. In this simulation, no slew requests which would stop the trigger are issued.

## IV. TRIGGER IMPLEMENTATIONS

The trigger algorithms have been implemented in C++ as a set of libraries, relying only on the external library FFTW [7] for the FFT computations and GSL [8] for the fits. They have been compiled in a standalone program, the **Scientific Software Model** (SSM), running on Linux machines.

The input to the SSM is a large file, containing the time-ordered simulated photons detected (pixel, energy, time), including background photons modulated by the Earth passages and source photons from simulated GRBs. The SSM walks through this file, simulating time-cycles for both trigger algorithms, and outputs GRB alert sequences. Fig. 8 shows an example of the algorithm running on the SSM.

In the **Data Processing Model** (DPM) of the UTS (see our simultaneously submitted paper II – Data Processing Model and Benchmarks [4]), the same trigger libraries have been cross-compiled for its AT697F (Leon-2) processor (ensuring that exactly the same trigger algorithms are running on both SSM and DPM), integrated with the RTEMS application, and interfaced with the data pre-processing FPGA which receives photons in real-time from an injector hardware, reading the same simulated photon files as the SSM.

The DPM permits to validate the UTS hardware-software system, in particular the CPU margin with the algorithms running. The SSM, being able to process photon files with much higher speed than real-time, is used to estimate the number of GRBs detected and optimize the algorithms.

## V. SIMULATION AND RESULTS

As input to the SSM, a dynamic simulation, based on individual photons with associated time, has been developed, using GRBs detected by the BATSE detector onboard the Compton-GRO mission operating from 1991 to 2000, which detected GRBs above 25 keV (and triggered above 50 keV). We use the complete BATSE database with 1889 GRBs [9] for which the fluences (in erg/cm$^2$) are available on 4 energy bands (25-50, 50-100, 100-300, >300 keV) as well as the durations and the light-curves in a 64-ms binning.

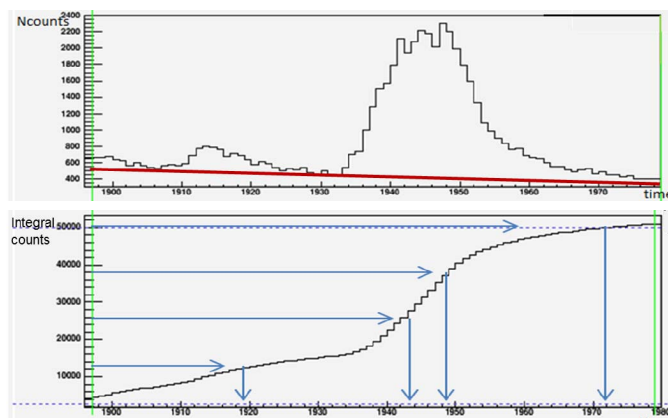

Fig. 9 – (top) Example light-curve with background fit. (bottom) The cumulative time-distribution permits to derive a time for as many photons as necessary. Their time-distribution will follow the light curve.

For each of those BATSE GRBs we build a file containing time-ordered photons with energy, such as expected to be impinging on ECLAIRs:

(a) Using the two lowest BATSE energy-bands (below Epeak), the photon-spectrum is modeled by a power-law $N(E)=A\,E^{-\lambda}$ with index $\lambda$. Doing so, the mean value of all $\lambda$ is overestimated by a factor 1.4 compared to the mean low-energy photon index of the 300 brightest GRBs, for which detailed spectral information is available [9]. Using a corrected power-law index $\lambda/1.4$ (reducing the number of photons at low energy), the spectrum of each BATSE GRB is extrapolated into the ECLAIRs energy range (down to 4 keV), and the number of photons expected in the ECLAIRs energy band (in case of an on-axis GRB) is determined, resulting in a list of photons with energy associated.

(b) To associate a time to each photon, the BATSE 64-ms light-curve of each GRB is used. First the background level is fit and subtracted. Then a cumulative time-distribution is produced (integral number of photons on the GRB duration vs. time), from which a uniform random-number generator allows the association of a time-value to each photon (Fig. 9).

We also generate 500 files with background photons expected to be detected by ECLAIRs, covering half an orbit each (45 min). For each one, an Earth transit through the field-of view is chosen at a random moment in a one-year-long minute-by-minute simulation of the spacecraft pointing evolution, as expected by the mission pointing strategy. Individual photons which follow the CXB energy spectrum [11] are placed isotropically inside the field-of-view. Using a Poisson time distribution, only photons not obscured by the Earth at their given time are accepted. Those photons are projected by ray-tracing through an ECLAIRs geometrical simulation (with mask open fraction of 40% and mask pixel size of 5.4 mm), which takes into account the detector inefficiency and mask transparency estimated by a GEANT-4 Monte-Carlo simulation [12] of ECLAIRs. Furthermore a spatially flat low-level instrumental background is added.

For each of the BATSE GRB photon files, a random background file is chosen. The GRB is placed at a random time inside the background file with an offset chosen between 5 to 25 min, to allow to the trigger algorithm sufficient time to start and process and a random isotropic position is attributed to the GRB inside the field-of-view. As long as the GRB happens to be obscured by the Earth, the choice of the GRB time and position is repeated. Once the GRB position is chosen, the GRB photons are projected through the ECLAIRs simulation onto the detection plane.

Both resulting photon lists, for the GRB and the background, are merged in time and run through the SSM, which implements the trigger algorithms as described previously (with an imaging SNR threshold of 6.5 σ, resulting in zero false alerts), and analyses with both trigger algorithms such a half-orbit file in about a minute on a standard Linux machine.

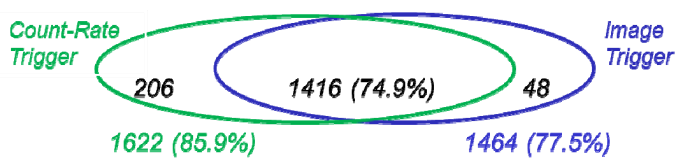

Fig. 10 – Result of the simulated BATSE GRB dataset detection by both UTS trigger algorithms.

The result of the simulation is that 88.4% of the BATSE GRBs are detected, 85.9% by the Count-Rate Trigger and 77.5% by the Image Trigger (Fig. 10). This fraction represents the product of the detector and the trigger efficiencies, most of the undetected bursts being short BATSE-GRBs, difficult to be detected by ECLAIRs, which needs a minimum number of counts (about 200) for the coded mask imaging being able to localize a source.

From this fraction one can estimate the number of BATSE-type GRBs per year to be triggered on by ECLAIRs/UTS. Starting from the figure of 666 BATSE-GRBs/yr on the whole sky [9], with a 85% duty cycle for a detector active outside the South Atlantic Anomaly, a field-of-view covering 13% of the sky, among which 70% is un-obscured by Earth, and using the detection fraction of 88.4%, one expects 45 BATSE-GRBs/yr.

Estimates from the HETE-II spacecraft, with a 2 keV low-energy threshold (similar to the one of ECLAIRs), show that 35% of its detected GRBs are X-ray rich GRBs and X-ray flashes [13], non-detected by BATSE. Therefore, thanks to its low energy threshold, ECLAIRs will detect 54% more GRBs than the standard BATSE-type GRBs. We thus expect 69 GRBs/yr to be triggered-on by the presented ECLAIRs configuration and trigger algorithms. This result will be confirmed in forthcoming work, based on the GRB datasets of Swift and Fermi-GBM, which benefit form a lower trigger threshold than BATSE, closer to the ECLAIRs one.

ACKNOWLEDGMENTS

The technical developments presented are co-financed by the CEA (Commissariat à l'Energie Atomique et aux Energies Alternatives) and the CNES (French Space Agency).